\documentclass[a4paper,12pt]{article}

\usepackage[utf8]{inputenc}
\usepackage[T1]{fontenc}
\usepackage{amsmath,amssymb}
\usepackage{graphicx}
\usepackage{siunitx}
\usepackage{authblk}
\usepackage{hyperref}
\usepackage[margin=2.5cm]{geometry}
\usepackage{lmodern}
\usepackage{siunitx}
\usepackage{graphicx}
\usepackage{tcolorbox}
\usepackage{caption}
\usepackage{floatrow}
\tcbuselibrary{listings, breakable}

\usepackage[numbers,sort&compress]{natbib} 

\title{\textbf{Thermography Equation: From Conceptual Relation to Quantitative Formulation via the Optogeometric Factor}}

\author[1]{Jan Sova}
\author[1]{Marie Kolaříková}
\affil[1]{Czech Technical University in Prague, Faculty of Mechanical Engineering}
\date{\today}

\begin{document}

\maketitle

\begin{abstract}
This article presents a methodological transition from the conceptually 
formulated thermographic measurement equation (the so-called thermography equation) 
to its quantitative form, expressed through the optogeometric factor 
\(F_{\mathrm{opg}}\). This factor directly links the known radiant flux 
from the measured surface to the energy captured by a single pixel of a thermal camera. 
The formulation is based on geometric–optical relations between the scene and the detector 
and is applicable to both scene-based and sensor-based approaches. 
The resulting expression provides a general framework for quantitative thermography.

\end{abstract}

\addcontentsline{toc}{section}{Abstract}


{
  \setlength{\parskip}{0pt}
  \tableofcontents
  \vspace{0.5\baselineskip}
  \hrule
}

\section{Introduction}

\label{sec:introduction}
The thermography equation is a fundamental relationship in infrared thermography~\cite{hs2025}.  
It relates the thermal radiation emitted by a measured scene to the radiant flux received by a single pixel of a thermal imaging camera through its optical system.  
In its general form, the equation accounts for three distinct components of the radiant flux:

\begin{enumerate}
  
\item \textbf{Thermal radiation emitted by the surface of the observed object}, determined primarily by its temperature and emissivity~\cite{Metal2023Nemec}. An ideal blackbody emits radiation according to Planck’s law; real surfaces emit less efficiently, with the reduction described by their emissivity \(\varepsilon\), which ranges from 0 to 1.

    \item \textbf{Thermal radiation reflected from the surface}, originating from surrounding objects or the environment. If the surface is not a perfect blackbody (its emissivity is less than 1), part of the incident radiation is reflected rather than emitted, contributing to the signal detected by the camera.

    \item \textbf{Thermal radiation from the atmosphere}, which is partly emitted by the atmospheric gases and partly results from the attenuation of radiation coming from the object’s surface. The atmosphere thus acts both as an additional source of radiation and as a medium that weakens the signal on its way to the camera.

\end{enumerate}

In some cases, such as measurements taken through an infrared (IR) window, the transmittance of the window must also be taken into account. It acts as a multiplicative factor that reduces all components of the radiant flux, further influencing the overall radiometric balance.

Historically, the thermography equation has often been presented in a \emph{conceptual} form, as shown in Section~\ref{sec:conceptual}. A \emph{quantitative formulation} of this equation, derived directly from physical and geometrical principles, was presented in~\cite{hs2025} with the original aim of correcting non-perpendicular infrared measurements in resistance spot welding~\cite{ForejtovaAutomotive, kolarik2014}, as well as, for instance, in UAV-based monitoring~\cite{Cukor2019}. However, it soon became evident that the resulting formulation has much broader applicability.

Section~\ref{sec:quantitative} introduces a \textbf{quantitative form of the thermography equation, representing a new concept in infrared thermography}. This formulation is expressed using the so-called optogeometric factor~\cite{opg2025}, which captures the purely geometrical relationship between the collecting area of the optical system and the solid angle subtended by a single detector pixel.

\section{Physical principles of thermal radiation}

This section outlines the physical principles of thermal radiation that underpin thermographic measurements and the thermography equation. It reviews the basic laws of emission, reflection, transmission, and atmospheric effects to provide a common framework for the conceptual and quantitative formulations presented later.

\subsection{Thermal radiation of bodies}

Thermal radiation, along with heat conduction, is one of the two fundamental mechanisms\footnote{Heat transfer is commonly divided into four modes: conduction, convection, radiation, and phase change (e.g., evaporation or condensation). However, only conduction and radiation are fundamental physical mechanisms.} of heat transfer. 
It occurs through the emission of electromagnetic radiation in the form of discrete quanta of energy, namely photons. 
As the surface temperature of an object increases, the intensity of its emitted radiation also increases. 
The spectral distribution of thermal radiation is described by Planck's law of blackbody radiation~\cite{vollmer2010}:
\begin{equation}
L_\lambda(T) = \frac{2 h c^2}{\lambda^5} \cdot 
\frac{1}{\exp\!\left(\frac{h c}{\lambda k_{\mathrm{B}} T}\right) - 1},
\label{eq:planck}
\end{equation}
where:
\begin{itemize}
  \item \( L_\lambda(T)\;[\si{\watt\per\meter\squared\per\steradian\per\meter}] \) is the spectral radiance at temperature \( T \) and wavelength \( \lambda \),
  \item \( h = \SI{6.62607015e-34}{\joule\second} \) is Planck’s constant,
  \item \( c = \SI{2.99792458e8}{\meter\per\second} \) is the speed of light in vacuum,
  \item \( k_{\mathrm{B}} = \SI{1.380649e-23}{\joule\per\kelvin} \) is Boltzmann’s constant,
  \item \( \lambda \) is the wavelength [\si{\meter}],
  \item \( T \) is the absolute temperature of the blackbody [\si{\kelvin}].
\end{itemize}

The total hemispherical radiant exitance \( M_B(T) \) of an ideal blackbody is obtained by integrating Planck’s law over all wavelengths and over the entire hemisphere of emission~\cite{vollmer2010}:
\begin{equation}
M_B(T) = \int_{0}^{\infty} M_\lambda(T) \,\mathrm{d}\lambda
= \int_{0}^{\infty} \!\left[ \int_{\Omega_\mathrm{hem}} L_\lambda(T) \cos\theta \, \mathrm{d}\Omega \right] \mathrm{d}\lambda .
\end{equation}
For isotropic (Lambertian) blackbody emission, the directional integration yields \(\pi L_\lambda(T)\), so that
\begin{equation}
M_B(T) = \pi \int_{0}^{\infty} L_\lambda(T) \, \mathrm{d}\lambda
= \sigma T^4 ,
\label{eq:SBlaw_black}
\end{equation}
where the Stefan–Boltzmann constant is
\begin{equation}
\sigma = \frac{2\pi^5 k_{\mathrm{B}}^4}{15 c^2 h^3}
= \SI{5.670374419e-8}{\watt\per\meter\squared\per\kelvin\tothe{4}} .
\label{eq:sigma}
\end{equation}

For a gray body with total hemispherical emissivity
\( \varepsilon \in [0,1] \), assumed constant across wavelengths and directions, 
the radiant exitance takes the form of
\begin{equation}
M = \varepsilon \, \sigma T^4 ,
\label{eq:graybody_exitance}
\end{equation}
while its spectral behavior is illustrated in Fig.~\ref{fig:sedeteleso}.

\begin{equation}
M_G(T) = \varepsilon\, \sigma T^4 .
\label{eq:graybody}
\end{equation}
Finally, the total radiant flux \( \Phi_{\text{obj,ref}}\;[\si{\watt}] \) 
emitted and reflected by the surface area of the measured object 
\( A\;[\si{\meter\squared}] \) is obtained as

\begin{equation}
\Phi_{\text{obj,ref}} = A \cdot M,
\label{eq:basic_radiant_flux}
\end{equation}
where \( M\;[\si{\watt\per\meter\squared}] \) is the surface exitance 
(including both emission and reflection).

where Eq.~\eqref{eq:basic_radiant_flux} expresses the fundamental relation
between radiant exitance $M$ and total radiant flux $\Phi_{\text{total}}$
emitted by a surface of area $A$.

\subsection{Radiative properties and kirchhoff’s laws}
\label{sec:radiative_props}

When electromagnetic radiation strikes the surface of a material, three basic interactions can occur:
\begin{itemize}
  \item a portion is \textbf{reflected} from the surface,
  \item a portion is \textbf{absorbed} within the material,
  \item and a portion is \textbf{transmitted} through the material.
\end{itemize}

These effects are quantified by dimensionless coefficients, each defined as the ratio of the corresponding radiant flux \( \Phi \) to the total incident flux \( \Phi_{\text{total}} \):
\begin{equation}
\rho = \frac{\Phi_{\text{ref}}}{\Phi_{\text{total}}}, 
\qquad
\alpha = \frac{\Phi_{\text{abs}}}{\Phi_{\text{total}}}, 
\qquad
\tau = \frac{\Phi_{\text{trans}}}{\Phi_{\text{total}}},
\label{eq:fundamental_definitions}
\end{equation}
where:
\begin{itemize}
  \item \( \rho\;[\si{-}]\) — reflectance,
  \item \( \alpha\;[\si{-}]\) — absorptance,
  \item \( \tau\;[\si{-}]\) — transmittance,
  \item \( \Phi_{\text{ref}}\;[\si{\watt}]\), \( \Phi_{\text{abs}}\;[\si{\watt}]\), \( \Phi_{\text{trans}}\;[\si{\watt}] \) — reflected, absorbed, and transmitted radiant fluxes,
  \item \( \Phi_{\text{total}}\;[\si{\watt}] \) — total incident radiant flux.
\end{itemize}

By conservation of energy the coefficients satisfy
\begin{equation}
\alpha(\lambda,T) + \rho(\lambda,T) + \tau(\lambda,T) = 1.
\label{eq:kirchhoff_sum}
\end{equation}
For \textbf{opaque} (non-transmitting) materials, \( \tau(\lambda,T)=0 \), hence
\begin{equation}
\alpha(\lambda,T) + \rho(\lambda,T) = 1.
\label{eq:opaque}
\end{equation}

\paragraph{Kirchhoff’s law at thermal equilibrium.}
At thermodynamic equilibrium the spectral emissivity equals the spectral absorptance:
\begin{equation}
\varepsilon(\lambda, T) = \alpha(\lambda, T).
\label{eq:kirchhoff_eq}
\end{equation}

Combining Eqs.~\eqref{eq:opaque} and \eqref{eq:kirchhoff_eq} for opaque materials gives the reflectance–emissivity relation:
\begin{equation}
\rho(\lambda, T) = 1 - \varepsilon(\lambda, T).
\label{eq:reflectance_opaque}
\end{equation}

In many practical applications, the emissivity 
\( \varepsilon \) is commonly approximated as constant with respect to both 
wavelength and temperature. Although this is an approximation, it is often 
sufficient within calibration accuracy and the uncertainties of material 
properties relevant to the measurement task. Under this assumption, the 
broadband form of Kirchhoff’s law simplifies to
\begin{equation}
\label{eq:reflectance_eps}
\rho = 1 - \varepsilon .
\end{equation}

Thus, for any real, non-transmitting surface, the reflectance is given by Eq.~\eqref{eq:reflectance_eps} — the form directly used in the conceptual thermography model (Eq.~\ref{eq:conceptual_thermography}).
In other words, any reduction in emissivity corresponds directly to an increase in the fraction of incident radiation that is reflected.

Here, \textbf{emissivity} is defined as the ratio of the radiant power emitted by a real surface to that emitted by an ideal blackbody at the same temperature and wavelength; it is a material- and wavelength-dependent measure of how effectively a surface emits thermal radiation.

\begin{figure}[H]
    \centering
    \includegraphics[width=0.9\textwidth]{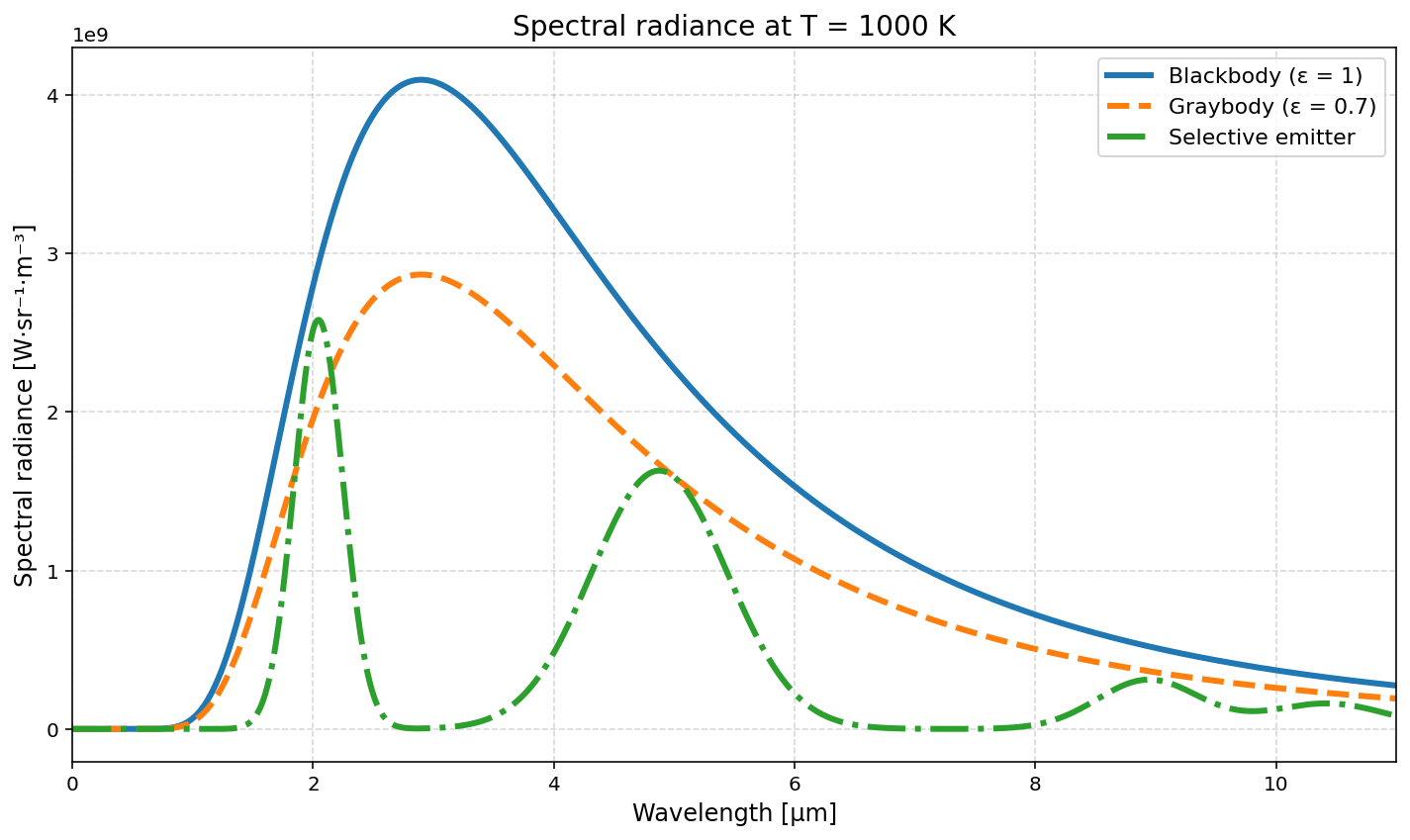}
    \caption{Spectral radiance at $T = \SI{1000}{\kelvin}$ for different emitter models.
The blackbody ($\varepsilon = 1$) follows Planck's law (solid line).
The graybody with constant emissivity $\varepsilon = 0.7$ (dashed line) has the same spectral shape but reduced magnitude.
The selective emitter (dash--dot line) has an emissivity modeled as a sum of Gaussian peaks, resulting in pronounced enhancement or suppression in specific spectral regions.}
    \label{fig:sedeteleso}
\end{figure}

Figure~\ref{fig:sedeteleso} illustrates emissivity concepts for three idealized radiators:
\begin{itemize}
    \item \textbf{Blackbody} — ideal emitter with \( \varepsilon(\lambda)=1 \) for all \( \lambda \); its spectral radiance follows Planck’s law exactly.
    \item \textbf{Gray body} — non-ideal emitter with constant \( \varepsilon(\lambda)<1 \) over all \( \lambda \); its spectrum is a uniform scaling of the blackbody spectrum.
    \item \textbf{Selective radiator} — emitter with wavelength-dependent \( \varepsilon(\lambda) \), often showing peaks/troughs due to material-specific absorption/emission bands.
\end{itemize}

\section{Thermography equation: conceptual relation}

The relation presented in this section corresponds to the traditional, 
conceptual form of the thermography equation 
\cite{vollmer2010,rogalski2002,minkina2009infrared}. 
It is essentially a radiative balance model that has been widely used for 
interpretative purposes, but it does not account for the geometric 
coupling between the observed surface and the detector pixel. 
In particular, it omits the optogeometric factor 
$F_{\mathrm{opg}}$ introduced in~\cite{hs2025}, 
which enables a full quantitative description of pixel-level measurements. 
Here, the conceptual relation is included for completeness and to 
facilitate the transition to the more rigorous formulation developed later.

\subsection{Formulation of the conceptual relation}

\begin{figure}[H]
    \centering
    \includegraphics[width=0.9\textwidth]{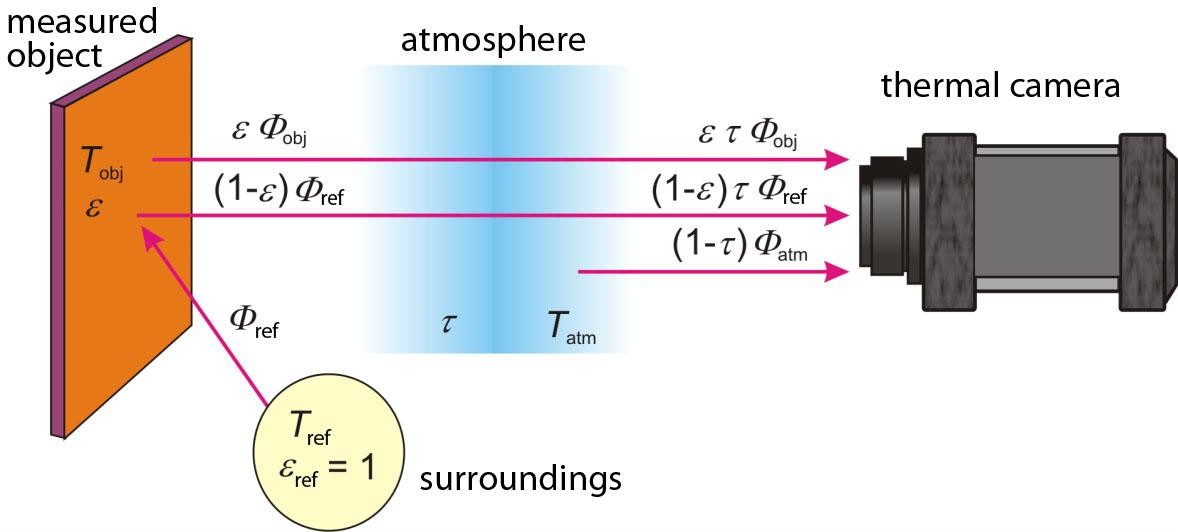}
    \caption{Conceptual model of radiative heat transfer in infrared thermography.
The diagram shows the three main sources contributing to the radiant flux $\Phi_{\text{total}}$ received by a single pixel of a thermal camera:
(i) thermal self-emission of the observed surface,
(ii) reflection of surrounding radiation from the surface (relevant when $\varepsilon < 1$), and
(iii) atmospheric emission along the optical path, combined with its attenuation of the other contributions.
This historical, balance-type representation does not account for geometric coupling between scene and detector. In modern quantitative models, such as the formulation in~\cite{hs2025}, this is resolved by introducing the optogeometric factor $F_{\mathrm{opg}}$ that links the projected scene area to the pixel footprint.}
    \label{fig:conceptual_model}
\end{figure}

In the classical conceptual approach to infrared thermography 
(illustrated in Fig.~\ref{fig:conceptual_model}), 
one considers a thermal camera observing the surface of an object whose 
temperature is to be determined. The radiant flux 
$\Phi_{\text{obj}}\;[\si{\watt}]$ emitted by the object surface and arriving at a detector pixel 
is described by Planck’s radiation law, or by its Stefan–Boltzmann approximation 
when spectral detail is not required. 
In addition to this self-emission, the pixel also receives the 
\emph{reflected flux} $\Phi_{\text{ref}}$, which originates from thermal radiation of the 
surroundings and is partially reflected by the surface when $\varepsilon < 1$. 
Finally, the \emph{atmospheric flux} $\Phi_{\text{atm}}$ contributes due to the emission 
of the atmosphere along the optical path, while at the same time attenuating the other 
two contributions. 

This formulation serves as a radiative balance model; 
it does not yet include the geometric coupling between scene and detector, 
which is addressed in the quantitative model introduced later in~\cite{hs2025}.

Since the measured surface (outside idealized conditions) always has an emissivity 
$\varepsilon\;[\si{-}]$ less than one ($\varepsilon < 1$), 
a fraction of the incident thermal radiation is reflected from the surface. 
As derived in Eq.~\eqref{eq:reflectance_eps}, 
the reflectance (that is, the reflection coefficient) of a material that does not transmit 
thermal radiation — only absorbs and reflects it — is given by $(1 - \varepsilon)$.

\label{sec:conceptual}
\begin{equation}
\Phi_{\text{total}} = \tau \cdot \left[ \varepsilon \ \Phi_{\text{obj}} + (1 - \varepsilon) \Phi_{\text{ref}} \right]   + (1-\tau) \Phi_{\text{atm}},
\label{eq:conceptual_thermography}
\end{equation}

where

\begin{itemize}
  \item $\Phi_{\text{total}}\;[\si{\watt}]$ is the total radiant flux reaching the camera pixel,
  \item $\tau\;[\si{-}]$ is the transmittance of the atmosphere or IR window,
  \item $\varepsilon\;[\si{-}]$ is the emissivity of the object's surface,
  \item $\Phi_{\text{obj}}\;[\si{\watt}]$ is the radiant flux emitted by the object surface,
  \item $\Phi_{\text{ref}}\;[\si{\watt}]$ is the radiant flux reflected from the object surface,
  \item $\Phi_{\text{atm}}\;[\si{\watt}]$ is the radiant flux emitted by the atmosphere toward the camera.
\end{itemize}

\subsection{Limitations of the conceptual equation}

The thermography equation can be expressed by substituting the radiant flux terms 
using the Stefan–Boltzmann law. This leads to a temperature-based formulation of the 
thermography equation, which is commonly found in the literature~\cite{vollmer2010,rogalski2002,minkina2009infrared}:

For clarity of explanation, we now neglect atmospheric effects and consider only the 
self-emission of the observed surface and the reflected radiation from its surroundings. 
The resulting radiant flux incident on a single detector pixel can then be expressed as

\begin{equation}
\Phi_{\text{obj,ref}} =
\left[
  \varepsilon \, \sigma T_{\text{obj}}^{4} +
  (1 - \varepsilon) \, \sigma T_{\text{ref}}^{4}
\right] \cdot A_{x},
\label{eq:SB_thermography_Ax}
\end{equation}

where $A_{x}\;[\si{m^{2}}]$ is an effective constant that converts surface exitance 
($\si{W\,m^{-2}}$) into radiant power at the pixel level ($\si{W}$), 
in direct analogy to the general radiometric relation given in Eq.~\eqref{eq:basic_radiant_flux}.

Importantly, the factor $A_{x}$ introduced in Eq.~\eqref{eq:SB_thermography_Ax} cannot be derived from the conceptual relation itself, which highlights the limitation of the balance-type model. The conceptual formulation does not provide any mechanism for determining the geometric coupling between the observed surface and the detector. In~\cite{hs2025}, this factor appeared in its reduced form, 

\begin{equation}
A_{x} \equiv \bar{F}_{\mathrm{opg}} = \frac{F_{\mathrm{opg}}}{\pi},
\label{eq:Fopg_reduced}
\end{equation}

as part of the quantitative derivation of the thermography equation. It was later formally established as a standalone physical quantity, the \emph{optogeometric factor}, in~\cite{opg2025}, where its general definition and broader applicability to imaging systems were presented.

\begin{tcolorbox}[
  colback=red!5,
  colframe=red!40,
  boxrule=0.5pt,
  title=Summary remark,
  coltitle=black
]
\textbf{Although the temperature-based formulation of the thermography equation 
(Eq.~\eqref{eq:SB_thermography_Ax}) is frequently used in the literature, 
it remains incomplete.} In fact, it is often written even without the reduced 
optogeometric factor $A_{x} \equiv \bar{F}_{\mathrm{opg}}$, which means that the 
resulting expression does not preserve the correct physical units: the radiant flux 
is expressed as if it were a surface exitance. The factor $A_{x}$, converting exitance 
($\si{W\,m^{-2}}$) into radiant power at the pixel level ($\si{W}$), cannot be obtained 
from the conceptual relation itself. A consistent quantitative formulation of the 
thermography equation, explicitly including this factor, was first provided in~\cite{hs2025}, 
and the optogeometric factor was later established as a standalone physical quantity with 
broader applicability in~\cite{opg2025}. Nevertheless, the simplified temperature-based 
form is still often employed as if it provided a complete quantitative description, which 
may lead to systematic misinterpretations in thermographic analysis.

\end{tcolorbox}

\section{Thermography Equation: Quantitative Relation}
\label{sec:quantitative}

The quantitative thermography equation is formulated using the optogeometric factor. 
Its derivation was presented in~\cite{hs2025}, while the radiometric definition of 
$F_{\mathrm{opg}}$ and its sensor-based form $F_{\mathrm{opg,s}}$, together with the 
reduced variants $\bar{F}_{\mathrm{opg}}$, $\bar{F}_{\mathrm{opg,s}}$ and the 
approximated relations $\tilde{F}_{\mathrm{opg}}$, $\tilde{F}_{\mathrm{opg,s}}$, 
as well as their reduced counterparts $\tilde{\bar{F}}_{\mathrm{opg}}$, 
$\tilde{\bar{F}}_{\mathrm{opg,s}}$, were given in~\cite{opg2025}.

\subsection{Optogeometric factor}
\label{sec:optogeometric_factor}

The \emph{optogeometric factor} is a purely geometrical quantity that links 
the radiant exitance of the observed surface (\si{W\,m^{-2}}) to the radiant flux 
received by a single detector pixel (\si{W}). It captures the joint effect of the 
collecting aperture of the optical system and the solid angle associated with the 
pixel footprint. In its general form~\cite{opg2025} it is defined as
\begin{equation}
F_{\mathrm{opg}} \equiv 
\iint_{A} \iint_{\Omega} 
\cos\theta \,\mathrm{d}\Omega \,\mathrm{d}A,
\label{eq:Fopg_def}
\end{equation}
which in the paraxial approximation reduces to
\begin{equation}
F_{\mathrm{opg}} \approx A \,\Omega,
\label{eq:Fopg_simple}
\end{equation}
where $A\;[\si{m^2}]$ is an effective collecting area and 
$\Omega\;[\si{sr}]$ the corresponding solid angle. 
The units of $F_{\mathrm{opg}}$ are \si{m^2 sr}.

where $A_{\mathrm{ap}}\;[\si{m^2}]$ is the effective area of the entrance pupil and 
$\Omega_{\mathrm{pix}}\;[\si{sr}]$ is the solid angle of the scene corresponding to 
one pixel. The units of $F_{\mathrm{opg}}$ are \si{m^2 sr}.

For convenience, a \emph{reduced form} of the factor is often used,
\begin{equation}
\bar{F}_{\mathrm{opg}} = \frac{F_{\mathrm{opg}}}{\pi},
\label{eq:Fopg_bar}
\end{equation}
with units of \si{m^2}. This reduced factor appeared in~\cite{hs2025} under the symbol 
$A_{x}$ as part of the first quantitative derivation of the thermography equation. 
It can be directly interpreted as the effective area that converts surface exitance 
into radiant power at the pixel level.

In addition to the scene-based definition, a sensor-based formulation 
$F_{\mathrm{opg,s}}$ was introduced in~\cite{opg2025}, relating the same geometric 
coupling to sensor parameters (focal length, pixel pitch, $f/\#$). 
Both $F_{\mathrm{opg}}$ and $F_{\mathrm{opg,s}}$ admit simplified expressions in 
the paraxial approximation, denoted by $\tilde{F}_{\mathrm{opg}}$ and 
$\tilde{F}_{\mathrm{opg,s}}$, respectively. When both the reduction and approximation 
are applied simultaneously, the notation $\tilde{\bar{F}}_{\mathrm{opg}}$ and 
$\tilde{\bar{F}}_{\mathrm{opg,s}}$ is used.

This hierarchy of definitions establishes the optogeometric factor as a standalone 
physical quantity~\cite{opg2025}, bridging scene-based and sensor-based approaches 
and enabling a consistent quantitative formulation of the thermography equation.

In practice, the most commonly employed forms are the \emph{approximated reduced factors}, 
which act as effective pixel-level constants in the thermography equation. 

Two practical parameterizations are used; they are mutually consistent under the paraxial thin-lens mapping $\varphi_{\mathrm{iFOV}}\!\approx a/f$ and $f\#=f/D$, but they are not strictly equivalent outside these assumptions

\begin{equation}
\tilde{\bar{F}}_{\mathrm{opg}}^{(D,\varphi)} 
= \frac{1}{4}\,\big(D \, \varphi_{\mathrm{iFOV}}\big)^{2},
\label{eq:Fopg_reduced_Dphi}
\end{equation}

\begin{equation}
\tilde{\bar{F}}_{\mathrm{opg,s}}^{(a,f\#)} 
= \frac{1}{4}\,\Big(\frac{a}{f\#}\Big)^{2},
\label{eq:Fopg_reduced_af}
\end{equation}

where $D$ is the entrance pupil diameter, $\varphi_{\mathrm{iFOV}}$ is the instantaneous 
field of view of a pixel, $a$ is the pixel pitch, and $f\#$ is the f-number of the optics. 
These two formulations are consistent representations of the same reduced factor only within the paraxial, no-vignetting regime (small angles, $\,\varphi_{\mathrm{iFOV}}\!\approx a/f$, $\,f\#=f/D$). Outside these conditions (finite object distance, field curvature, vignetting, non-telecentric imaging, non-square pixels) they are distinct approximations and need not coincide.

\begin{table}[H]
\centering
\caption{Notation summary for the optogeometric factor. Emphasis is placed on the two
approximated reduced forms (last two rows), which are used as pixel-level constants
in the thermography equation.}
\label{tab:Fopg_notation}
\begin{tabular}{lllc}
\hline
\textbf{Symbol} & \textbf{Name} & \textbf{Expression (if applicable)} & \textbf{Units} \\
\hline
$F_{\mathrm{opg}}$ & scene-based factor & $A_{\mathrm{ap}}\,\Omega_{\mathrm{pix}}$ & \si{m^{2}\,sr} \\
$\bar{F}_{\mathrm{opg}}$ & reduced scene-based factor & $\displaystyle \frac{F_{\mathrm{opg}}}{\pi}$ & \si{m^{2}} \\
$F_{\mathrm{opg,s}}$ & sensor-based factor & (sensor formulation, see~\cite{opg2025}) & \si{m^{2}\,sr} \\
$\bar{F}_{\mathrm{opg,s}}$ & reduced sensor-based factor & $\displaystyle \frac{F_{\mathrm{opg,s}}}{\pi}$ & \si{m^{2}} \\
\hline
$\tilde{F}_{\mathrm{opg}}$ & approximated scene-based & $\pi\,\tilde{\bar{F}}_{\mathrm{opg}}$ & \si{m^{2}\,sr} \\
$\tilde{F}_{\mathrm{opg,s}}$ & approximated sensor-based & $\pi\,\tilde{\bar{F}}_{\mathrm{opg,s}}$ & \si{m^{2}\,sr} \\
\hline
\multicolumn{4}{l}{\textit{Approximated reduced (used as constants in the thermography equation)}}\\
$\tilde{\bar{F}}_{\mathrm{opg}}^{(D,\varphi)}$ & approx.\ reduced (aperture–angle) 
& $\displaystyle \frac{1}{4}\,\big(D\,\varphi_{\mathrm{iFOV}}\big)^{2}$ & \si{m^{2}} \\
$\tilde{\bar{F}}_{\mathrm{opg,s}}^{(a,f\#)}$ & approx.\ reduced (pixel–$f\#$)
& $\displaystyle \frac{1}{4}\,\Big(\frac{a}{f\#}\Big)^{2}$ & \si{m^{2}} \\
\hline
\end{tabular}

\vspace{0.5em}
\footnotesize
$A_{\mathrm{ap}}$ – effective entrance pupil area; 
$\Omega_{\mathrm{pix}}$ – pixel solid angle; 
$D$ – entrance pupil diameter; 
$\varphi_{\mathrm{iFOV}}$ – pixel field angle (in radians); 
$a$ – pixel pitch; 
$f\#$ – f-number. 
The reduced forms divide by $\pi$ (cf.\ $\bar{F}=F/\pi$), while tildes denote paraxial approximations.
\end{table}

\subsection{Quantitative thermography equation}
\label{sec:thermography_equation}

In the simplest case, considering only the object’s own emission and 
the reflected ambient radiation while neglecting atmospheric influence, 
the radiant flux received by a single pixel is expressed as

\begin{equation}
\Phi_{\text{obj,ref}}^{(\text{pix})} \;=\;
\Big[\;
  \varepsilon\,\sigma T_{\text{obj}}^{4}
  \;+\;
  (1-\varepsilon)\,\sigma T_{\text{ref}}^{4}
\;\Big]\;\frac{F_{\mathrm{opg}}}{\pi},
\label{eq:thermo_Fopg_general}
\end{equation}

where $F_{\mathrm{opg}}\,[\si{m^{2}sr}]$ is the scene-based optogeometric factor. 
The division by $\pi$ reflects the Lambertian relation between exitance and radiance. 
Here the subscript ``obj,ref'' denotes that the flux originates from the object’s emission 
and the reflected ambient radiation, while the superscript $(\text{pix})$ indicates 
that it is expressed at the level of a single detector pixel.

\medskip
\noindent
For practical calculations, it is often convenient to use the 
\emph{reduced approximated forms} of the factor. In the paraxial regime, 
the following expressions are obtained:

\begin{tcolorbox}[
  colframe=black,        
  colback=white,         
  sharp corners,         
  boxrule=1pt,           
  title={Thermography equations formulated at the pixel level}
]
\begin{align}
\Phi_{\text{obj,ref}}^{(\text{pix})}
&\;\approx\;
\Big[\;
  \varepsilon\,\sigma T_{\text{obj}}^{4}
  \;+\;
  (1-\varepsilon)\,\sigma T_{\text{ref}}^{4}
\;\Big]\;
\tilde{\bar{F}}_{\mathrm{opg}}^{(D,\varphi)},
\label{eq:thermo_reduced_Dphi}
\\[6pt]
\Phi_{\text{obj,ref}}^{(\text{pix})}
&\;\approx\;
\Big[\;
  \varepsilon\,\sigma T_{\text{obj}}^{4}
  \;+\;
  (1-\varepsilon)\,\sigma T_{\text{ref}}^{4}
\;\Big]\;
\tilde{\bar{F}}_{\mathrm{opg,s}}^{(a,f\#)},
\label{eq:thermo_reduced_af}
\end{align}
\end{tcolorbox}

with

\begin{tcolorbox}[
  colframe=black,
  colback=white,
  sharp corners,
  boxrule=1pt,
  title={Pixel-level approximations of the optogeometric factor}
]
\[
\tilde{\bar{F}}_{\mathrm{opg}}^{(D,\varphi)} = 
\frac{1}{4}\,\big(D\,\varphi_{\mathrm{iFOV}}\big)^{2},
\qquad
\tilde{\bar{F}}_{\mathrm{opg,s}}^{(a,f\#)} = 
\frac{1}{4}\,\left(\frac{a}{f\#}\right)^{2}.
\]
\end{tcolorbox}

\medskip
\noindent
Equations~\eqref{eq:thermo_reduced_Dphi} and 
\eqref{eq:thermo_reduced_af} represent two alternative approximations 
of the same reduced factor, which coincide only in the paraxial, 
no-vignetting regime. Within this framework, the factor provides a 
direct link between surface exitance and detected radiant power, 
forming the basis of the quantitative thermography equation first 
derived in~\cite{hs2025} and later generalized in~\cite{opg2025}.

\subsection{Quantitative thermography equation incorporating angular dependence of emissivity}

Surface emissivity is generally not isotropic but decreases with increasing incidence
angle of thermal radiation. A commonly used empirical model for this angular
dependence~\cite{Nicodemus70, canopyemisssivity} is

\begin{equation}
\varepsilon(\alpha) = \varepsilon_0 \cos^n(\alpha),
\label{eq:emissivity_model}
\end{equation}

where $\alpha$ is the angle measured from the surface normal, $\varepsilon_0$ is the
normal (nadir) emissivity, and $n$ is an empirical exponent depending on the material.
This model provides a simple but effective way to approximate the directional
behavior of real surfaces, with typical ranges:

\begin{itemize}
  \item $n \approx 0$ for nearly Lambertian materials (e.g., matte paints, ceramic surfaces),
  \item $n \approx 1$ to $n \approx 2$ for common building materials (e.g., plaster, concrete, brickwork),
  \item $n > 3$ for shiny or metallic surfaces (e.g., aluminum, copper, chrome).
\end{itemize}

For some materials, the exponent $n$ may also vary with measurement wavelength or
surface treatment.

\begin{figure}[H]
  \vspace{2pt}\hrule height 0.4pt\relax\vspace{4pt}
    \centering
    \includegraphics[width=1\textwidth]{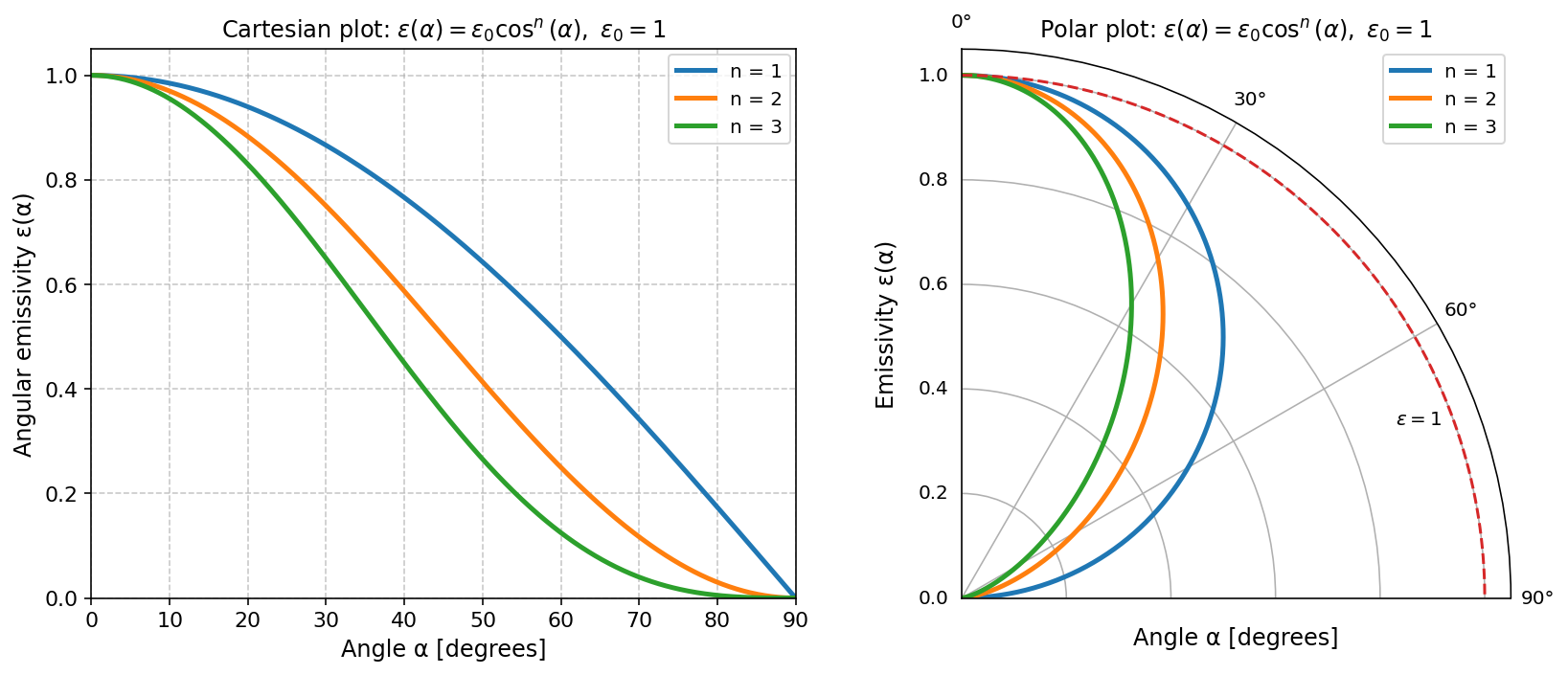}
    \caption{Angular dependence of emissivity modeled as
    $\varepsilon(\alpha)=\varepsilon_0\cos^n(\alpha)$
    for three exponents ($n=1,2,3$). The incidence angle
    $\alpha$ is measured from the surface normal ($0^\circ$–$90^\circ$).
    Larger $n$ values yield steeper angular decay, showing how emissivity
    decreases with angle and how the exponent controls directional sensitivity.
    The left panel shows the model in Cartesian coordinates
    ($\varepsilon$ vs. $\alpha$), while the right panel shows the same in polar
    coordinates ($\varepsilon$ plotted radially against $\alpha$).
    For clarity, $\varepsilon_0=1$ is assumed.}
    \label{fig:cosine_emisivity}
  \vspace{4pt}\hrule height 0.4pt\relax\vspace{2pt}
\end{figure}

By substituting Eq.~\eqref{eq:emissivity_model} into the quantitative thermography
framework, the detected radiant flux can be expressed as

\begin{equation}
\Phi_{\text{pix}}(\alpha) =
\left[
  \varepsilon_0 \cos^n(\alpha) \, \sigma T^4 +
  \bigl(1 - \varepsilon_0 \cos^n(\alpha)\bigr) \, \sigma T_{\text{ref}}^4
\right] \cdot \bar{F}_{\mathrm{opg}}^{(D,\varphi)},
\label{eq:thermo_emissivity_angle}
\end{equation}

where $\tilde{\bar{F}}_{\mathrm{opg}}^{(D,\varphi)}$ is the reduced approximate
optogeometric factor in the aperture–angle formulation
[Eq.~\eqref{eq:Fopg_reduced_Dphi}].
 This equation generalizes the quantitative
thermography model by explicitly incorporating directional emissivity effects,
as illustrated in Fig.~\ref{fig:cosine_emisivity}.

\section{Conclusion}
\label{sec:conclusion}

This article has presented a systematic transition from the traditional
conceptual thermography equation to its quantitative formulation,
achieved through the explicit introduction of the optogeometric factor.
Unlike the conceptual balance model, which cannot determine the geometric
coupling between scene and detector, the quantitative equation provides
a physically consistent link between surface exitance and the radiant
power received by a single pixel.

We have summarized the hierarchy of definitions of the optogeometric
factor, including its scene-based, sensor-based, reduced, and
approximated forms. In practical use, the approximated reduced variants
serve as effective pixel-level constants, enabling straightforward
quantitative evaluation of thermographic measurements.

Finally, the formulation was extended by incorporating an empirical
cosine-based model of the angular dependence of emissivity, which allows
directional surface properties to be taken into account. This extension
further improves the realism of the quantitative thermography equation
and illustrates its flexibility for applications beyond the simplified
Lambertian assumption.

Overall, the framework presented here replaces the conceptual approach
with a unified quantitative model that can serve as a basis for
calibration, error analysis, and the development of advanced correction
procedures in infrared thermography.


\end{document}